\documentclass[twocolumn,showpacs,preprintnumbers]{revtex4-1}

\usepackage{graphicx}
\usepackage{dcolumn}
\usepackage{bm}
\usepackage{color}
\usepackage{textcomp}

\begin{document}

\title{Radiofrequency spectroscopy of $^6$Li $p$-wave molecules:\\towards photoemission spectroscopy of a $p$-wave superfluid}

\author{R.\,A.\,W{\kern -0.1em}. Maier}
\author{C. Marzok}
\author{C. Zimmermann}

\affiliation{Physikalisches Institut, Eberhard Karls Universit\"at
T\"ubingen, \\ Auf der Morgenstelle 14, D-72076 T\"ubingen,
Germany}

\author{Ph.\,W{\kern -0.1em}. Courteille}

\affiliation{Instituto de F\'isica de S\~ao Carlos, Universidade
de S\~ao Paulo, Caixa Postal 369, S\~ao Carlos-SP 13560-970,
Brazil}

\date{\today}

\begin{abstract} Understanding superfluidity with higher order partial waves is
crucial for the understanding of high-$T_c$ superconductivity. For
the realization of a superfluid with anisotropic order parameter,
spin-polarized fermionic lithium atoms with strong $p$-wave
interaction are the most promising candidates to date. We apply
rf-spectroscopy techniques that do not suffer from severe
final-state effects \cite{Perali08} with the goal to perform
photoemission spectroscopy on a strongly interacting $p$-wave
Fermi gas similar to that recently applied for s-wave interactions
\cite{Stewart08}. Radiofrequency spectra of both quasibound
$p$-wave molecules and free atoms in the vicinity of the $p$-wave
Feshbach resonance located at 159.15\,G \cite{Schunck05} are
presented. The observed relative tunings of the molecular and
atomic signals in the spectra with magnetic field confirm earlier
measurements realized with direct rf-association \cite{Fuchs08}.
Furthermore, evidence of bound molecule production using adiabatic
ramps is shown. A scheme to observe anisotropic superfluid gaps,
the most direct proof of $p$-wave superfluidity, with 1d-optical
lattices is proposed.
\end{abstract}


%  34.50.-s    Scattering of atoms and molecules
%  67.85.Lm    Degenerate Fermi gases
%  74.20.Rp    Pairing symmetries (other than s-wave)

\maketitle

%--------------------------------------------------------------
%%%%%%%% Introduction %%%%%%%%%%
%--------------------------------------------------------------

In the last few years, great progress has been made in the study
of strongly interacting gases with $s$-wave symmetry, particularly
in the regime of crossover physics between quantum phases like the
BEC-BCS crossover \cite{Altmeyer07}, but also in the context of
ultracold molecule production \cite{Jochim03,Regal03}. At the same
time, only a relatively limited amount of experimental data was
collected in the vicinity of higher orbital angular momentum
Feshbach resonances, even though the anisotropy of $p$-wave
interactions allows for richer order parameters than in the case
of $s$-wave interactions. While for $s$-wave systems it is known
that observables change smoothly when the interaction strength is
varied across a Feshbach resonance \cite{Bourdel04,Bartenstein04},
theoretical calculations predict both smooth crossover physics
\cite{Ohashi05} as well as various quantum phase transitions
\cite{Botelho05, Iskin06} across a $p$-wave resonance, e.g. the
transition from a $p_x$-wave superfluid to a time-reversal
breaking $p_x+ip_y$-wave superfluid \cite{Gurarie05}. Such quantum
phase transitions could be observed directly by means of
rf-spectroscopy \cite{Iskin08}. Depending on the interaction
strength, the superfluid phase is either stable ("weak") or
unstable ("strong") \cite{Levinsen07}, while in optical lattices,
even more new and exotic superfluid phases are expected
\cite{Iskin05}. By studying higher partial wave superfluidity with
ultracold atoms, insight into the pairing effects of other
condensed matter systems could also be gained. For instance,
$^3$He \cite{Vollhardt90} and the more exotic heavy-fermion
superconductors like UPt$_3$ \cite{Stewart84} show $p$-wave
symmetry of the order parameter while high-$T_c$ superconductors
are known to exhibit $d$-wave symmetry \cite{Damascelli03}.

Moreover, only few data on ultracold $p$-wave molecules are
available. First $p$-wave molecules made from ultracold $^{40}$K
have been studied at JILA and only very short lifetimes on the
order of 1\,ms were observed for the different $m_l$ projections.
Above threshold, the molecular lifetime is limited by tunneling
through the centrifugal barrier. The binding energies were found
to scale linearly with detuning from the resonance
\cite{Gaebler07}. This was used to determine the relative magnetic
moment of $^6$Li $p$-wave molecules by means of rf-spectroscopy
\cite{Fuchs08}. Studies of the collisional stability of $^6$Li
$p$-wave molecular samples have also been performed
\cite{Inada08}.

Recently, photoemission spectroscopy methods were applied for the
first time to study a strongly interacting spin mixture of
$^{40}$K atoms \cite{Stewart08}. By coupling atoms from the
strongly interacting states into a third, only weakly interacting
state, severe final state effects that have strong influence on
the shape of the spectra \cite{Perali08}, can be avoided. Due to
the negligible rf-photon recoil, the measured momentum
distribution of the outcoupled atoms can be used to directly
determine the dispersion relation in the strong coupling regime. A
BCS-like dispersion curve with an energy offset resembling a
superfluid gap was observed \cite{Stewart08}.

In this brief report, we apply the final state effect free
rf-spectroscopy technique to a strongly interacting gas of $^6$Li
atoms in their $\left|1\right>$ state close to the well-known
$p$-wave Feshbach resonance located at about 159.15\,G
\cite{Schunck05}. We detect both atoms and quasibound molecules
with positive binding energy by coupling them to the weakly
interacting $\left|2\right>$ state ($\left|F=1/2,m_F=-1/2\right>$
at low fields). Comparing the shift of the molecular feature
relative to the atomic peak for different magnetic fields, we
confirm the observations made in \cite{Fuchs08} as to the relative
magnetic moments of the molecular and atomic states
$\left|1\right>$.

Furthermore, we produce bound molecules by means of adiabatic
magnetic field ramps similar to those used in \cite{Inada08}.
After application of a purification pulse that is resonant to the
atomic transition while the photon recoil energy $E_{\rm
rec}\approx h\times 74\,$kHz is below the binding energy of the
molecules, we observe a small number of bound molecules. Lifetime
measurements of the molecules in the presence of residual unbound
atoms are compatible with inelastic coefficient determinations in
\cite{Inada08}.

%--------------------------------------------------------------
%%%%%%%% Apparatus and experimental Procedure %%%%%%%%%%
%--------------------------------------------------------------
The cooling procedure for producing an ultracold mixture of $^6$Li
and $^{87}$Rb atoms has been detailed in a previous publication
\cite{Silber05}. In summary, we simultaneously collect both
species in a magneto-optical trap, polarize their spins through
optical pumping, trap them magnetically and transfer them via
several intermediate magnetic traps into a Joffe-Pritchard type
trap. Here they are stored in a stretched state mixture of the
Zeeman hyperfine states $^6$Li\,$|F=3/2,\,m_F=3/2\rangle$ and
$^{87}$Rb\,$|F=2,\,m_F=2\rangle$. In the rotationally symmetric
Joffe-Pritchard trap, which has trapping frequencies of
$(\omega_r, \omega_z)/2\pi =(762, 190)\,\text{Hz}$ for $^6$Li and
a magnetic field offset of $3.5\,$G, the $^{87}$Rb is selectively
cooled by hyperfine state changing microwave evaporation at
6.8\,GHz. The $^6$Li sympathetically cools through interspecies
thermalization. At a common temperature of about 3\,\textmu K with
$3\times10^6$ $^{87}$Rb atoms and about $1\times10^5$ $^6$Li atoms
we stop the evaporation and move the atoms into a second
Joffe-Pritchard geometry, which is located in the center between
the trap coils. We then load the atoms into a horizontally
oriented crossed beam optical dipole trap generated by an
intensity-stabilized fiber laser running at 1064\,nm (IPG
YLR-10LP). At 2.9\,W and 3.2\,W of laser power in each arm and
equal beam waists of 58\,\textmu m, the trap depths are about
$130\,$\textmu K ($^{87}$Rb) and $45\,$\textmu K ($^{6}$Li) with
trap frequencies of $\tilde\omega_{\rm
Rb}/2\pi\approx(610\times440\times440)^{1/3}$\,Hz ($^{87}$Rb) and
$\tilde\omega_{\rm
Li}/2\pi\approx(1.3\times0.9\times0.9)^{1/3}$\,kHz ($^{6}$Li).
After loading into the optical trap, the $^6$Li has a temperature
of about 9\,\textmu K. The trap coils are then used to generate a
homogeneous magnetic bias field. We perform a radio frequency
Landau-Zener transition at about 228\,MHz to transfer the $^6$Li
atoms into the absolute groundstate
$\left|1\right>:=\left|F=1/2,m_F=1/2\right>$. The $^{87}$Rb atoms
are removed from the trap by means of a resonant optical pulse of
1\,ms duration on the cycling transition
$5\,^2S_{1/2}\left|F=2\right>\rightarrow5\,^2P_{3/2}\left|F'=3\right>$
also used for imaging. We then perform two kinds of experiments:
in the first experiment, we ramp the magnetic field to a specific
value that we want to probe with rf-spectroscopy. By varying the
rf-frequency, we measure a spectrum by counting the atoms that are
transferred into the state $\left|2\right>$. In a second
experiment we study molecule production via adiabatic magnetic
field ramps across the resonance. We find a very limited
efficiency for this method.

%--------------------------------------------------------------
%%%%%%%% Results %%%%%%%%%%%%%%%%
%--------------------------------------------------------------
%\textbf{- Molecules in rf-spectra}
%--------------------------------------------------------------
\begin{figure}[ht]
\centerline{\scalebox{.43}{\includegraphics{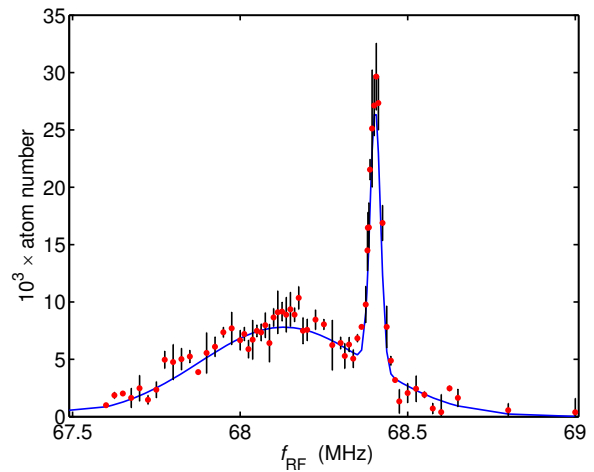}}}
\caption{(Color online) RF-spectrum measured at a magnetic
detuning of $-60$\,mG relative to the maximum of three-body loss
determined by atom loss measurements. The broad molecular peak is
shifted 270\,kHz to the red from the narrow atomic peak. This
positive binding energy corresponds to a magnetic detuning of
$B-B_{\rm res}=+85\,$mG. The molecules are therefore quasibound
due to the $p$-wave centrifugal barrier and the maximum of the
three body loss is shifted relative to the actual Feshbach
resonance at $B_{\rm res}$. The observed shift of $+145\,$mG is
compatible with theoretical predictions due to finite temperature
effects \cite{Chevy05}. Each datapoint (red dots) is the average
of up to seven single experimental cycles, the corresponding
standard deviation is plotted as black error bar. The blue curve
is a fit with two gaussian functions to determine the frequencies
of maximum atomic and molecular signal.} \label{fig:spectrum}
\end{figure}

In Fig. \ref{fig:spectrum} we present a typical rf-spectrum of the
atom-molecule mixture close to the 159.15\,G $p$-wave Feshbach
resonance \cite{Schunck05}. The spectrum is measured by coupling
atoms/molecules into the atomic state $\left|2\right>$ with a
2\,ms long rf-pulse at about 68\,MHz after we ramp the magnetic
field to a value about 60\,mG below the maximum of atomic
three-body loss associated with the Feshbach resonance. Following
the application of the rf-pulse we quickly ($<50$\,\textmu s) ramp
the magnetic field to a value approximately $150$\,mG higher to an
imaging field held constant for all measurements, turn off the
optical trap and image the state $\left|2\right>$ using the
cycling transition
$\left|2\right>=2\,^2S_{1/2}\left|J=1/2,\,m_J=-1/2,\,I=1,\,m_I=0\right>\rightarrow
2\,^2P_{3/2}\left|J'=3/2,\,m_J'=-3/2,\,I'=1,\,m_I'=0\right>$ after
a certain amount of time-of-flight. We observe two distinct
features: a narrow peak which corresponds to the atomic transition
and a broad feature that corresponds to $p$-wave molecules that
have a positive binding energy of approximately $270$\,kHz
relative to the atomic threshold. This binding energy corresponds
to about +85\,mG magnetic field detuning relative to the crossing
point of the molecular and the atomic state \cite{Fuchs08}. From
the positive binding energy it is clear, that these quasibound
molecules are stabilized by the centrifugal barrier of the
$p$-wave collision channel. Comparing spectra at different
magnetic fields, we observe a shift between the molecular peaks of
$(115\pm 30)$\,kHz for two different magnetic detunings to the
Feshbach resonance while the shift of the atomic peaks amounts to
$(2.7\pm0.3)$\,kHz. The atomic shift can be used to determine the
relative magnetic field shift between the two spectra: the
rf-transition has a magnetic tuning that is approximately linear
with magnetic field in the vicinity of the Feshbach resonance with
a tuning of 87\,kHz/G. The magnetic tuning of the molecular state
relative to the atomic state $\left|1\right>$ is known to be
2.354\,MHz/G \cite{Fuchs08}. Taking the Zeeman shift of the
rf-coupled state $\left|2\right>$ into account, the molecular
state moves with 2.267\,MHz/G relative to state $\left|2\right>$.
The magnetic field shift is $\Delta B=(31\pm4)$\,mG as calculated
from the atomic peak shifts, resulting in a predicted molecular
peak shift of $(70\pm8)$\,kHz. This confirms the determination of
the relative magnetic moments of the molecular state and the state
$\left|1\right>$ performed in \cite{Fuchs08} within the accuracy
of our resolution of the molecular spectra reasonably well.

We note, that even though the spectrum in Fig. \ref{fig:spectrum}
was measured at a detuning of $-60$\,mG from the maximum of the
loss peak when resolving the Feshbach resonance with inelastic
atom loss measurements, we clearly observe quasibound molecules
with positive binding energy relative to the atomic state
$\left|1\right>$ corresponding to a positive detuning of
approximately +85\,mG. Therefore, the maximum of inelastic atom
loss does not correspond to the magnetic field of atom-molecule
state degeneracy, i.e. the actual Feshbach resonance. This is due
to the finite collision energy of the $^6$Li atoms at a
temperature of about 9\,\textmu K. A theo\-retical estimate for
the finite collision energy shift for this resonance was given as
$+0.1\,$G for $T=10\,$\textmu K in \cite{Chevy05} which is in
reasonable agreement with our observation of about $+ 145\,$mG.
Further, varying dynamical Stark effects between the molecular
state, the strongly interacting state $\left|1\right>$ and the
weakly interacting state $\left|2\right>$ could lead to additional
shifts due to the optical trapping potential \cite{Chin00}, but
have not been investigated in detail.

%--------------------------------------------------------------
%\textbf{- Molecules through adiabatic ramps}
%--------------------------------------------------------------

\begin{figure}[ht]
\centerline{\scalebox{.5}{\includegraphics{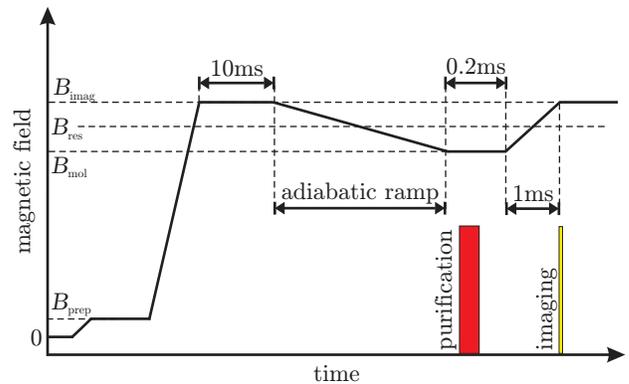}}}
\caption{(Color online) Time sequence for the generation of
$p$-wave molecules from a spin-polarized Fermi gas. After
preparation of the state $\left|1\right>$ at a field $B_{\rm
prep}$ we quickly (ca. 1\,ms) ramp to $B_{\rm imag}>B_{\rm res}$,
where the molecular energy is above the atomic threshold. The
field is held for 10\,ms to allow possible overshooting of the
field to decay. Then, a slow sweep towards a lower field $B_{\rm
mol}<B_{\rm res}$ is performed with variable ramp speed,
adiabatically transforming atoms into molecules. Next, a
25\,\textmu s purification pulse is applied in a small time window
of 0.2\,ms, removing all unpaired atoms. Finally, the molecules
are adiabatically dissociated back into atoms in 1\,ms for imaging
at the initial field $B_{\rm imag}$. Due to the small amount of
molecules produced, imaging is performed \textit{in situ} for
higher S/N ratio.} \label{fig:timesequence}
\end{figure}

\begin{figure}[ht]
\centerline{\scalebox{.43}{\includegraphics{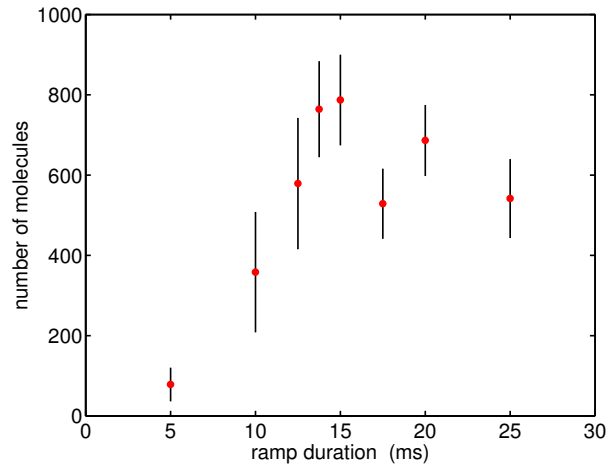}}}  % plotted against ramp duration
\caption{Number of bound $p$-wave molecules ($E_{\rm mol}<0$)
produced through adiabatic ramping of the magnetic field across
the Feshbach resonance with ramp width $\Delta B=305\,$mG. Only
small amounts of bound molecules were produced, probably due to
small phase space overlap at 9\,\textmu K. The error bars
correspond to one standard deviation for five to eight
measurements per data point.} \label{fig:molecules}
\end{figure}

We have also studied the production of $p$-wave molecules at the
low magnetic field side of the Feshbach resonance, where they form
bound molecules with negative binding energy with respect to the
atomic asymptote. For this we employ a method similar to that used
by the Tokyo group \cite{Inada08}. After preparation of the
desired state $\left|1\right>$ at a low magnetic field $B_{\rm
prep}$ we quickly ramp across the Feshbach resonance with
$dB/dt=71.7$\,G/ms to a value $B_{\rm imag}$ approximately 150\,mG
above $B_{\rm res}$. We hold $B_{\rm imag}$ for 10\,ms to allow
for decay of a possible overshooting of the current control
circuit. We then slowly ramp down to a magnetic field of $B_{\rm
mol}<B_{\rm res}$ with ramp width $\Delta B=B_{\rm imag}-B_{\rm
mol}=305\,$mG with varying ramp speed. We then remove the
remaining unbound atoms in a small time window of 0.2\,ms by means
of a 25\,\textmu s optical pulse resonant to the
$\left|1\right>=2\,^2S_{1/2}\left|J=1/2,m_J=-1/2,I=1,m_I=1\right>\rightarrow
2\,^2P_{3/2}\left|J'=3/2,m_J'=-3/2,I'=1,m_I'=1\right>$ transition
to prepare a pure molecular sample before we ramp back to $B_{\rm
imag}$ above the resonance in 1\,ms. This second ramp dissociates
the molecules that were not affected by the resonant light back
into free atoms. We image these atoms with absorption imaging
using the same resonant light that was used for purification. The
number of molecules produced in this way is shown in Fig.
\ref{fig:molecules}. Due to the low atom-molecule conversion
efficiency, we perform the imaging \textit{in situ} to increase
the S/N ratio. Only about 3\,\% of all atoms are transformed into
molecules in contrast to observations in \cite{Inada08}, where up
to 15\,\% were converted. This is probably due to the higher
temperature of 9\,\textmu K here compared to 1\,\textmu K in
\cite{Inada08} and the corresponding reduced phase space overlap.
We have studied the lifetime of the molecular sample produced in
this way in the presence of the remaining unbound atoms by holding
the atom/molecule mixture for a variable amount of time after the
adiabatic ramp and before the purification pulse. The value for
the inelastic atom/dimer-coefficient quoted in \cite{Inada08} as
$K_{dd}=2.4^{+0.5}_{-0.3}\times10^{-11}\,{\rm cm}^3/$s for
$\left|1\right>$-$\left|1\right>$ molecules is consistent with our
observations.

%--------------------------------------------------------------
%%%%%%%% Summary %%%%%%%%%%
%--------------------------------------------------------------
In conclusion, we have successfully applied recent final state
effect free rf-spectroscopy to an ultracold system of
spin-polarized $^6$Li atoms with strong anisotropic interaction
through a $p$-wave Feshbach resonance. We present spectra of
molecular and atomic features in the vicinity of the resonance and
confirm earlier findings as to the binding energy of quasibound
$p$-wave molecules. This technique now applied to a $p$-wave
system can be readily extended to study $p$-wave superfluidity by
mapping the momentum distribution and its anisotropy in order to
measure the dispersion relation of the ultracold gas in different
directions similar to the case of $s$-wave symmetry
\cite{Stewart08}. The currently limited atom/molecule numbers and
the resulting S/N ratio in our time-of-flight images prevents a
successful conversion into three dimensional density distributions
as inverse Abel transformation is very sensitive to noise
\cite{Dribinski02}.

After the realization of $s$-wave superfluidity, $p$-wave
superfluidity will mark a further milestone in ultracold atom
experiments. Several techniques for studying superfluid properties
have been proposed for cold atom experiments
\cite{Bruun00,Iskin08}. Reaching the extremely low temperatures
needed to achieve superfluidity remains the major obstacle
\cite{Houbiers97}; the Fermi-Bose mixture $^6$Li-$^{87}$Rb
constitutes an important candidate for reaching deep Fermi
degeneracy through sympathetic cooling in specialized trap
geometries \cite{Onofrio03}. Furthermore, $T_c$ can be manipulated
by applying dc electric fields \cite{You99}, thus lowering the
demands as to the degree of quantum degeneracy required.

%--------------------------------------------------------------
%\textbf{- Proposal for experiments in 1d-lattice:}
%--------------------------------------------------------------
For future experiments we propose to use a one-dimensional optical
lattice that confines the atoms into an effectively
two-dimensional space ($\omega_z>\hbar k_F^2/2m$). The magnetic
bias field should be perpendicular to the 1d-planes of the optical
lattice, thereby defining the $m_l$-projections relative to that
plane. The direction of imaging should be chosen perpendicularly
to the lattice beams such that the images can reveal the $p$-wave
symmetry of the scattering process for specific $m_l$-projections.
Addressability of different $m_l$-projections is easier in
$^{40}$K as it has a large magnetic dipole-dipole shift of
$0.5$\,G between $m_l=0$ and $m_l=\pm1$
\cite{Ticknor04,Gaebler07}, while for $^6$Li this shift is
predicted to be only small \cite{Chevy05}. Dispersion curves of
the gas taken by measuring rf-spectra into a non-interacting final
state like $\left|2\right>$ with subsequent imaging of the final
state as in \cite{Stewart08} could be able to reveal the
anisotropic superfluid gap in the dispersion curves after inverse
Abel transformation. The quantum-phase transition along the
BEC-BCS crossover could also be detected using direct
rf-spectroscopy \cite{Iskin08}.

Future work is devoted to increasing the number of cold $^6$Li
atoms as well as increasing the degree of quantum degeneracy. An
extension of our experimental setup with an optical lattice will
enable us to polarize the $p$-wave collisions inside a plane, thus
making anisotropic effects visible in photoemission measurements
under improved experimental conditions.

This work has been supported by the Deutsche
Forschungsgemeinschaft (DFG). We are thankful for stimulating
discussions with A. Muramatsu.

\end{document}